\documentstyle[aps,prl,amstex,amssymb,epsfig,floats,preprint,tighten]{revtex}
\begin{document}
\draft

\preprint{BARI - TH 97-263}
\date{ March 1997}
\title{Variational method and duality in the 2D square Potts model}
\author{L. Angelini \cite{angelini}, M. Pellicoro \cite{pellicoro}, 
I. Sardella \cite{sardella}, M. Villani \cite{villani}}
\address{
Dipartimento di Fisica dell'Universit\`a di Bari,
70126 Bari, Italy\\
{\rm and}\\
Istituto Nazionale di Fisica Nucleare, Sezione di Bari,
70126 Bari, Italy
}
\maketitle
\begin{abstract}
The ferromagnetic q-state Potts model on a square lattice is analyzed, for 
$q>4$, through an elaborate version of the operatorial variational method. In 
the variational approach proposed in the paper, the duality relations are 
exactly satisfied, involving at a more fundamental level, a duality relationship 
between variational parameters. Besides some exact predictions, the approach is 
very effective in the numerical estimates over the whole range of temperature 
and can be systematically improved.
\end{abstract}

\section{Introduction}
\par We consider the ferromagnetic $q$-state Potts model \cite{Wu82} on a 
square 
lattice, limiting ourselves to the isotropic, nearest-neighbour (n.n.) case, 
with no external field. The hamiltonian of the model is given by 
\begin{equation}
{H \over kT} = - K \sum_{<ij>} \delta_{s_i s_j}, \label{1}
\end{equation}
where $s_i = 1,2,\dots,q$, $i$ denotes the lattice sites, $\delta _{s_is_j}$ 
is the Kronecker symbol and $K = J/kT$, with $J > 0$. We suppose that the 
lattice is made of $m$ rows and $n$ columns, and that free  boundary conditions 
are applied. Our interest will be in the limits $m,n \to \infty$.
\par For $q > 4$ the model has a temperature driven first-order phase 
transition  at $K_t=K_t(q)= \ln (1+\sqrt{q}) $. At $K_t$ the free energy $f$, 
the internal energies $u_o$ and $u_d$ of the ordered and disordered phase, and 
the latent heat 
are known exactly \cite{Wu82,Baxter82}. Still at $K_t$, more recently an 
analytic formula for the correlation length has been also derived 
\cite{BW93,KSZ89}.
\par Due to these exact results, the $2D$ Potts model has an important role in 
the 
study of the first order phase transitions \cite{Binder87}. As a matter of fact 
its properties have been studied through several methods, such as analysis of 
high 
or low-temperature series-expansions \cite{BEG94}, resummation of large $q$
expansions 
\cite{BLM93,BLM94,BLM96}, Monte Carlo simulations 
\cite{Billoire92,BNB93,JK94,JK95,JK95a}. As a rule, the discontinuous character 
of the first order phase 
transitions makes the numerical calculations in these cases more uncertain, 
with 
respect to analogous calculations in the case of second order phase 
transitions. 
Moreover we have that some aspects of the discontinuous transitions are not 
completely understood from a general, theoretical point of view.
\par On the other hand an accurate description of the Potts model transition, 
when 
the number of components $q$ is very large, is provided by the Mean Field 
Theory 
\cite{MS74}. Indeed , it has been proved that Mean Field Theory is exact in the 
limit $q \to +\infty$ \cite{PG80}. So it suggests itself that a reliable 
quantitative analysis of the properties of our model, at least for $q>4$, could 
be also obtained directly from a proper extension of this theory.
\par As a matter of fact, in this paper it is shown that considerable and 
systematic improvements of the Mean Field Theory can be obtained in the 
framework of the transfer matrix approach, through an elaborate version of the 
standard variational method for hermitian operators. This method is alternative 
to the variational Gibbs principle \cite{PG80,Huang87}, which provides the 
analytical basis of the mean field procedure or, more generally, of the Cluster 
variational method \cite{Morita72}.
\par The formal aspects of our approach are described in the next four 
sections. 
We will limit ourselves to $q>4$. In the last section we present the results of 
some numerical calculations and we add some concluding remarks.

\section{Operatorial variational method}
Let $Z_{m,n}(K)$ be the partition function of our model. We are interested in 
the free energy per site, $f(T)$, given by
\begin{equation*}
f(T)=-kT \phi(K),
\end{equation*}
where
\begin{equation}
\phi(K)=\lim_{m \to \infty} \lim_{n \to \infty} {1 \over {n m}} \log Z_{m,n}(K)
\label{2}
\end{equation}
Now, if $\sigma = (s_1,s_2,\dots,s_m)$ describes the spin configuration on a 
column, we can write $Z_{m,n}$ in the form
\begin{equation}
Z_{m,n}=\sum_{\sigma_1,\sigma_2,\dots,\sigma_n} \chi(\sigma_1) 
L(\sigma_1,\sigma_2) L(\sigma_2,\sigma_3) \dots L(\sigma_{n-1},\sigma_n) 
\chi(\sigma_n) = (\chi,L^{n-1} \chi), \label{3}
\end{equation}
where
\begin{equation}
\chi(\sigma)=e^{{K \over 2}\sum_{i=1}^{m-1} \delta_{s_i,s_{i+1}}}
\label{4}
\end{equation}
and
\begin{equation}
L(\sigma,\sigma')=\chi(\sigma)e^{K \sum_{i=1}^{m} 
\delta_{s_i,s'_i}}\chi(\sigma') \label{5}
\end{equation}
is the symmetrized transfer matrix of our model.
\par The matrix $L(\sigma,\sigma')$ is hermitian. Therefore we can apply the 
spectral theorem and we can take into account of the extremal properties of its 
eigenvalues. We note also that, for our $L(\sigma,\sigma')$, the 
Perron-Frobenius  (P.F.) theorem is valid \cite{CF72}. Furthermore we have that 
$L(\sigma,\sigma')$ is positive definite, that is, for any vector 
$\xi(\sigma)$, we obtain
\begin{equation}
(\xi,L\xi)=\sum_{\sigma,\sigma'} {\bar \xi}(\sigma) L(\sigma,\sigma') 
\xi(\sigma') \geq 0. 
\label{6}
\end{equation}
In fact we can write
\begin{equation*}
(\xi,L\xi)=\sum_{\sigma,\sigma'} {\bar f}(\sigma) \prod_{i=1}^m (1+(e^K -1) 
\delta_{s_i,s'_i}) f(\sigma'),
\end{equation*}
where
\begin{equation*}
f(\sigma)=\xi(\sigma) \chi(\sigma).
\end{equation*}
By expanding the product, we obtain a sum of $2^m$ terms, each giving a non 
negative contribution, due to $e^K -1 > 0$.
\par Let $\lambda_1(K;m)$ be the highest eigenvalue of $L(\sigma,\sigma')$. 
From 
\eqref{2} and \eqref{3} it follows that \cite{CF72}:
\begin{equation}
\phi(K) = \lim_{m \to \infty} {1 \over m} \log \lambda_1(K,m).
\label{7}
\end{equation}
Now, by considering the functional given by the Rayleigh-Ritz (R.R.) quotient
\begin{equation}
F[\xi] = {(\xi,L\xi) \over (\xi,\xi)},
\end{equation}
we have
\begin{equation*}
\sup_\xi F[\xi] = \lambda_1(K;m),
\end{equation*}
the $\sup$ being attained for $\xi(\sigma) = \psi_1(\sigma)$, where 
$\psi_1(\sigma)$ is the eigenvector of $L(\sigma,\sigma')$ corresponding to 
$\lambda_1(K;m)$. So, for any vector $\xi(\sigma)$, it results that
\begin{equation}
\lim_{m \to \infty} {1 \over m} \log {(\xi,L\xi) \over (\xi,\xi)} \le 
\phi(K)
\label{9}
\end{equation}
This inequality is the starting point of the operatorial variational method in 
Statistical Mechanics \cite{Munster74,Rujan79}. In this approach, the principal 
(or ground state) eigenvector $\psi_1(\sigma)$ is usually described through an 
ansatz $\tilde\psi(\sigma;\alpha,\beta,\dots)$, containing one or several 
unknown parameters which do not depend on $m$. Then, within the formal 
structure 
of the considered ansatz, the best estimate of $f(T)$ is obtained by fixing the 
parameters through
\begin{equation}
\sup_{\alpha,\beta,\dots} \lim_{m \to \infty} {1 \over m} \log 
{(\tilde\psi,L\tilde\psi) \over (\tilde\psi,\tilde\psi)}
\label{10}
\end{equation}
As a matter of fact, this approach, which is at the origin of the concept of 
the 
transfer matrix, is not really adequate to analyse the properties of a critical 
point. However the situation can be different in the case of first order 
transitions.
\par In the following, fixing the attention on the 2D square Potts model, we 
will discuss a further elaboration of the above variational method.
\par First of all, it is useful to bear in mind the meaning and the role of 
$\psi_1(\sigma)$. In our lattice made of $m$ rows, let us consider a strip 
$\Sigma_i$ of $i$ adjacent columns, having $n_r$ columns at its right and $n_l$ 
columns at its left. The reduced probability distribution (r.p.d.) 
$P_i = P_i(\sigma_1, \sigma_2, \dots, \sigma_i)$ of a spin configuration on
$\Sigma_i$ , obtained by averaging over all the configurations at the right and 
the left of $\Sigma_i$, is given, in the limit $n_r,n_l \to +\infty$, by
\begin{equation}
P_i(\sigma_1, \sigma_2, \dots, \sigma_n) = 
{ \psi_1(\sigma_1)L(\sigma_1, \sigma_2)L(\sigma_2, \sigma_3)
\dots L(\sigma_{i-1}, \sigma_i)\psi_1(\sigma_i) 
\over {(\psi_1,L^{i-1} \psi_1})} ,
\label{11}
\end{equation}
for $i \geq 2$, while in the case of one column, we have the r.p.d. $P_1 
(\sigma)$ through
\begin{equation}
P_1 (\sigma) = { \psi_1^2 (\sigma) \over {(\psi_1,\psi_1)}}
\label{12}
\end{equation}
So, an ansatz $\tilde\psi(\sigma)$ for $\psi_1(\sigma)$ has the role of a 
parametric description of the above averaging process leading to the r.p.d. 
$P_i$.
\par Now, very reliable and simple descriptions, having however a different 
formal content, can be obtained by considering separately the high and low 
temperature regions. This fact drives us to introduce for $\tilde\psi(\sigma)$ 
two formal structures, generally different, which we call 
$\tilde\psi_d(\sigma)$ 
and $\tilde\psi_o(\sigma)$ and to consider $\tilde\psi(\sigma)$ as made of two 
{\it charts} $(\tilde\psi_d(\sigma),\tilde\psi_o(\sigma))$, associated to the 
ordered and disordered phase, which should match, if possible, at the 
transition 
point $K_t$. The choice of the appropriate {\it chart} is demanded to the 
variational method. This point of view is a sligth generalization of the 
traditional approach, where a fixed formal structure for $\tilde\psi(\sigma)$ 
is 
considered, and only the parameters are varied. However it leads quite 
naturally 
to variational approximations to $\phi(K)$, in which duality 
\cite{Baxter82,Savit80} is exactly satisfied.

\section{One-parameter ansatz's}
In this section we introduce and discuss some one-parameter ansatz's of which 
we 
will make use in the variational approach based on equation \eqref{10}.
\par Let us fix the attention first on the disordered phase. For large $T$, due 
to the smallness of the correlation length, we expect that
\begin{equation}
P_1 (\sigma) \simeq {e^{K \sum_{i=1}^{m-1} \delta_{s_i,s_{i+1}}} \over 
(\chi,\chi)}
\label{13}
\end{equation}
Then, with regard to the disordered phase, we introduce the simplest one 
parameter ansatz for $\psi_1(\sigma)$ in the form 
\begin{equation}
\tilde\psi_d^{(1)}(\sigma)=e^{{A(K) \over 2} {\sum_{i=1}^{m-1} 
\delta_{s_i,s_{i+1}}}}, \qquad (A(K)>0)
\label{14}
\end{equation}
that is, for $K$ belonging to the disordered region, the result of the 
averaging 
process at the left and the right of a column, is represented simply by a 
renormalized ferromagnetic coupling $A(K)$ between n.n. spins, such that 
$A(K)\simeq K$ for small $K$.
\par This simplest ansatz can be improved in several ways. Since the averaging 
process produces new interactions among the spins of a column, we need, in 
principle, other parameters related to these new couplings. For example, a 
first 
correction to $\tilde\psi_d^{(1)}(\sigma)$ can be considered by taking into 
account also second-neighbour interactions. However, it is possible to stay 
with 
a one parameter ansatz and, at the same time, to improve $\tilde\psi_d^{(1)} 
(\sigma)$, by applying the transfer matrix $L$ to $\tilde\psi_d^{(1)}(\sigma)$. 
As a matter of fact, if $L$ is applied repeatedly to $\tilde\psi_d^{(1)} 
(\sigma)$, we reach at the end the principal eigenvector. Since $L$ is positive 
definite, we can consider, more generally, the action of $L^{1 \over 2}$. So, 
due to the simple structure of $\tilde\psi_d^{(1)}(\sigma)$, we will fix our 
attention on the sequence of one-parameter ansatz's
\begin{equation}
\tilde\psi_d^{(1)}(\sigma), \quad \tilde\psi_d^{(2)}(\sigma) = L^{1 \over 2} 
\tilde\psi_d^{(1)}(\sigma),\quad \dots \quad, \tilde\psi_d^{(\nu)}(\sigma) = 
L^{\nu \over 2} \tilde\psi_d^{(1)}(\sigma)
\label{15}
\end{equation}
The reliability of one approximate description of $\psi_1(\sigma)$ can be 
tested 
by considering the next approximation in the sequence.
\par Then we are led to the R.R. quotients
\begin{equation}
\rho_d^{(1)}(A,K)={{(\tilde\psi_d^{(1)}, L \tilde\psi_d^{(1)})} 
\over {(\tilde\psi_d^{(1)},\tilde\psi_d^{(1)})}} ,\qquad
\rho_d^{(2)}(A,K)={{(\tilde\psi_d^{(2)}, L \tilde\psi_d^{(2)})} 
\over {(\tilde\psi_d^{(2)},\tilde\psi_d^{(2)})}} , \quad \dots,
\label{16}
\end{equation}
which can be also written in the form
\begin{equation}
\rho_d^{(\nu)}(A,K) ={{(\tilde\psi_d^{(1)}, L^\nu \tilde\psi_d^{(1)})} \over 
{(\tilde\psi_d^{(1)},L^{\nu-1} \tilde\psi_d^{(1)})}}.
\label{17}
\end{equation}
We note that
\begin{equation}
Z_d^{(\nu)} (A,K) = (\tilde\psi_d^{(1)}, L^{\nu -1} 
\tilde\psi_d^{(1)})
\label{18}
\end{equation}
is, for $\nu>1$, the partition function of a Potts$(\nu;A)$ model on a strip 
$\Sigma_\nu$ of $m$ rows and $\nu$ columns, having the same couplings as the 
original model \eqref{1}, except on the first and last column, where there is 
the vertical n.n. coupling $(A(K)+K)/2$. This model provides an 
approximate description $\tilde P_d^{(\nu)}$ of the r.p.d. 
$P_\nu(\sigma_1,\sigma_2, \dots, \sigma_\nu)$.
\par We also remark that to Potts$(\nu;A)$ is associated a $q^\nu \times q^\nu$ 
transfer matrix $\ell_\nu(A,K)$, connecting two adjacent rows of $\Sigma_\nu$. 
We will call $\lambda_1^{(\nu)}(A,K)$ the highest eigenvalue of $\ell_\nu(A,K)$.
\par In each R.R. quotient we can fix the parameter $A(K)$, by considering the 
variational estimate \eqref{10}. We obtain then a sequence of estimates 
$\phi_d^{(\nu)}(K)$ of $\phi(K)$, given by
\begin{equation}
  \begin{split}
  \phi_d^{(\nu)}(K) &= \sup_A \lim_{m \to \infty} {1 \over m} \log 
                       {Z_d^{(\nu+1)}(A,K) \over Z_d^{(\nu)}(A,K)} \\
                    &= \sup_A \log {\lambda_1^{(\nu+1)}(A,K) \over              
                       \lambda_1^{(\nu)}(A,K)} \equiv \sup_A 
                       \phi_d^{(\nu)}(A,K) \qquad (\nu = 1,2,\dots)
   \end{split}
\label{19}
\end{equation}
where the $\sup$ is taken over values of $A$ which give ferromagnetic 
couplings, 
that is $A \geq 0$ for $\nu=1$, $A \geq -K$ for $\nu=2,3,\dots$.
\par It is useful to point out that, for $0 \leq A,K < +\infty$, the P.F. 
theorem 
is valid for the {\it reduced} transfer matrices of finite order 
$\ell_\nu(A,K)$, 
so that, in this range of the parameters, the Potts$(\nu;A)$ model has a unique 
{\it ground state}, as it happens in the thermodynamic limit for the original 
model, in the disordered phase.
\par Now, let us consider the ordered phase. For finite $m$, even if large, 
$\psi_1(\sigma)$ is a well defined, unique, vector; furthermore 
$P_1(\sigma)$ has the same symmetries of the hamiltonian \eqref{1}. However, in 
the limit $m \to \infty$, we have several phenomena as asymptotic degeneracy 
\cite{Baxter82} of the highest eigenvalue of $L$, related to the occurrence of 
$q$ ordered phases, macroscopic instabilities with respect to boundary 
perturbations or the existence of {\it spontaneous} long range order. So, an 
ansatz $\tilde\psi_o(\sigma)$ will be appropriate if these aspects can be 
correctly predicted. In particular, if we introduce {\it reduced} transfer 
matrices of finite order, as we did in the disordered case, we have to elude 
the P.F. theorem.
\par We show that the above requirements can be satisfied through an 
approximate 
description not of the result of the averaging process leading to
$\psi_1(\sigma)$, but of the process itself.
\par Let us consider a column $\Sigma$. We replace the $n_r$ columns, with $n_r 
\to \infty$, at the right of $\Sigma$, with one column $\Sigma''$ of Potts 
spins 
$\sigma_1'',\sigma_2'',\dots,\sigma_m''$, having an effective horizontal 
coupling 
$H (H>0)$ with the spins of $\Sigma$. Furthermore, adjacent spins $\sigma_i''$ 
are 
constrained to be in the same state, this last condition having the role to 
describe the ordered phase. The averaging process at the right of $\Sigma$ is 
then 
simulated by the summation over all configurations on $\Sigma''$. The same 
procedure is applied at the left of $\Sigma$, by introducing a column 
$\Sigma'$. 
So, we obtain the distribution
\begin{equation}
\tilde P_o(\sigma)= {1 \over \Gamma} \sum_{\sigma'} \sum_{\sigma''} 
\Pi(\sigma',\sigma) \Pi(\sigma'',\sigma) e^{K \sum_{i=1}^{m-1} 
\delta_{s_i,s_{i+1}}},
\label{20}
\end{equation}
where $\Gamma$ is a normalization constant and
\begin{equation}
\Pi(\sigma',\sigma)=e^{H \sum_{i=1}^{m} \delta_{s'_i,s_i}} \prod_{j=1}^{m-1} 
\delta_{s'_j,s'_{j+1}}.
\label{21}
\end{equation}
We are then led to the one-parameter ansatz $\tilde\psi_o^{(1)}(\sigma)$ in the 
ordered phase, given by
\begin{equation}
\tilde\psi_o^{(1)}(\sigma)=\left(\sum_{\sigma'} \Pi(\sigma',\sigma)\right) 
e^{{K \over 2} \sum_{i=1}^{m-1} \delta_{s_i,s_{i+1}}}
\label{22}
\end{equation}
We note that $\tilde P_o(\sigma)$ is a r.p.d. obtained from the distribution
\begin{equation}
\tilde P_o(\sigma',\sigma,\sigma'') = {1 \over \Gamma}{\Pi(\sigma',\sigma) 
\Pi(\sigma'',\sigma) 
e^{K \sum_{i=1}^{m-1} \delta_{s_i,s_{i+1}}}}
\label{23}
\end{equation}
involving the strip $\Sigma' \cup \Sigma \cup \Sigma''$. Now, to $\tilde 
P_o(\sigma',\sigma,\sigma'')$ is associated a $q^3 \times q^3$ transfer matrix 
$\ell=\ell(s'_i,s_i,s''_i|s'_{i+1},s_{i+1},s''_{i+1})$, having many matrix 
elements equal to zero, due to the factors $\prod_{i=1}^{m-1} 
\delta_{s'_i,s'_{i+1}}$, $\prod_{i=1}^{m-1} \delta_{s''_i,s''_{i+1}}$. This 
allows 
to elude the P.F. theorem. In fact, the highest eigenvalue of $\ell$ is $q$ 
times 
degenerate; we have that $\ell$ is the direct sum of the $q$ blocks made by the 
same $q \times q$ matrix,
\begin{equation*}
\ell(\mu) = \ell(s_i,s_{i+1};\mu) = \ell(\mu,s_i,\mu|\mu,s_{i+1},\mu) \qquad 
(\mu=1,2,\dots,q)
\end{equation*}
and of the $q(q-1)$ blocks
\begin{equation*}
\ell(\mu,\mu') =\ell(s_i,s_{i+1};\mu,\mu')=\ell(\mu,s_i,\mu'|\mu,s_{i+1},\mu') 
\qquad (\mu,\mu'=1,2,\dots,q \quad \mu \not= \mu')
\end{equation*}
made also by $q \times q$ coincident matrices, which differ from the previous 
ones. The highest eigenvalue of $\ell$ is contained in the blocks $\ell(\mu)$, 
which are associated to the $q$ ordered phases, while $\ell(\mu,\mu')$ are 
related 
to the occurrence of a surface tension between different phases. 
\par The problem of selecting a particular $\ell(\mu)$ is physically equivalent 
to 
apply an external field $h_\mu$ to the spin state $\mu$ , on the boundaries 
$\Sigma'$ and $\Sigma''$, to take the limit $m \to \infty$ and then to consider 
the limit $h_\mu \to 0$. Once this selection of a particular $\mu$ has been 
made, 
we see from \eqref{21} that the horizontal effective coupling $H$ gives rise to 
an 
external field $2H$, applied on $\Sigma$ to the spin state $\mu$, obtaining 
then 
the long range order relative to the $\mu$ phase. So, we see that our parameter 
$H$ has a role analogous to the effective field of the Mean Field Theory. By 
modulating through $H$ the rigid boundary conditions on $\Sigma'$ and 
$\Sigma''$, 
$\tilde\psi_o^{(1)}(\sigma)$ provides the simplest description of the phase 
having 
a temperature dependent order parameter.
\par As we did for the disordered phase, we can improve 
$\tilde\psi_o^{(1)}(\sigma)$ through the action of $L^{1 \over 2}$. So, in the 
ordered region, we have the sequence of one-parameter ansatz's for 
$\psi_1(\sigma)$
\begin{equation}
\tilde\psi_o^{(1)}(\sigma), \quad \tilde\psi_o^{(2)}(\sigma) = L^{1 \over 2} 
\tilde\psi_o^{(1)}(\sigma),\quad \dots 
\label{24}
\end{equation}
and the relative R.R. quotients
\begin{equation}
\rho_o^{(\nu)}(H,K) ={{(\tilde\psi_o^{(1)}, L^\nu \tilde\psi_o^{(1)})} \over 
{(\tilde\psi_o^{(1)},L^{\nu-1} \tilde\psi_o^{(1)})}} \qquad (\nu \geq 1)
\label{25}
\end{equation}
It is useful to point out that $\tilde\psi_o^{(1)}(\sigma)$ can be written in 
the form
\begin{equation}
\begin{split}
  \tilde\psi_o^{(1)}(\sigma) 
     &=\lim_{R \to +\infty} e^{-(m-1)R} \sum_{\sigma'} e^{R \sum_{i=1}^{m-1}
     \delta_{s'_i,s'_{i+1}} + H \sum_{i=1}^m \delta_{s'_i,s_i} + {K \over 2}
     \sum_{i=1}^{m-1}\delta_{s_i,s_{i+1}}} \\
     &\equiv \lim_{R \to +\infty} \tilde\psi_o^{(1)} (\sigma;R),
\end{split}
\label{26}
\end{equation}
and that
\begin{equation}
(\tilde\psi_o^{(1)}, L^{\nu -1} \tilde\psi_o^{(1)}) = Z_o^{(\nu+2)} (H,K) 
\qquad 
(\nu \geq 1)
\label{27}
\end{equation}
is the partition function of a Potts$(\nu+2;+\infty,H)$ model on a strip 
$\Sigma_{\nu+2}$ of $m$ rows and $\nu+2$ columns, with the same coupling as the 
original model \eqref{1}, except for the first and last column of sites, where 
we 
have a vertical n.n. coupling of infinite strength, and for the first and last 
column of rows where we have an horizontal n.n. coupling of strength $H$. This 
model provides an approximate description $\tilde P_o^{(\nu)}$ of the r.p.d. 
$P_\nu(\sigma_1,\sigma_2, \dots, \sigma_\nu)$ in the ordered phase.
\par To the Potts$(\nu+2;+\infty,H)$ is associated a $q^{\nu+2}\times 
q^{\nu+2}$ 
transfer matrix $\ell_{\nu+2}(+\infty,H,K)$ having properties analogous to that 
of the previous $\ell=\ell_3(+\infty,H,K)$. If we call 
$\lambda_1^{(\nu+2)}(+\infty,H,K)$ its highest eigenvalue, we obtain the 
sequence of estimates $\phi_o^{(\nu)}(K)$ of $\phi(K)$ in the form
\begin{equation}
\phi_o^{(\nu)}(K)=\sup_{H \geq 0} \log {\lambda_1^{(\nu+3)}(+\infty,H,K) \over 
\lambda_1^{(\nu+2)}(+\infty,H,K)} \equiv \sup_{H \geq 0} \phi_o^{(\nu)}(H,K) 
\qquad (\nu = 1,2,\dots)
\label{28}
\end{equation}
By taking a vector $\tilde\psi_d^{(i)}(\sigma)$ of the sequence \eqref{15} and 
a 
vector $\tilde\psi_o^{(j)}(\sigma)$ of the sequence \eqref{24}, we obtain an 
ansatz
\begin{equation}
\tilde\psi(\sigma) = (\tilde\psi_d^{(i)}(\sigma),\tilde\psi_o^{(j)}(\sigma)),
\label{29}
\end{equation}
with two {\it charts}, providing the estimates of $\phi(K)$, given by 
$\phi_d^{(i)}(K)$ and $\phi_o^{(j)}(K)$ respectively. These two functions are 
defined for any $K$. For each $K$, the choice between the two descriptions is 
made 
by a further application of the variational method, that is by considering
\begin{equation}
\max (\phi_d^{(i)}(K),\phi_o^{(j)}(K))
\label{30}
\end{equation}
We will say that two vectors $\tilde\psi_d^{(i)}(\sigma)$ and 
$\tilde\psi_o^{(j)}(\sigma)$ are exactly compatible if it happens that 
$\phi_d^{(i)}(K)$ and $\phi_o^{(j)}(K)$ match exactly at the transition point 
$K_t$. 
In the next section, we will show that such vectors exist.

\section{Duality}
A crucial property of the 2D Potts model is the duality relation which connects 
the partition function in the high and low-temperature regions. In terms of the 
free energy, it states that \cite{Baxter82}
\begin{equation}
\phi(K) = \log x^2 + \phi(K^*),
\label{31}
\end{equation}
where
\begin{equation*}
x={{e^K - 1} \over \sqrt{q}},
\end{equation*}
and $K^*$, the point {\it dual} of $K$, is defined by
\begin{equation*}
{{e^{K^*} - 1} \over \sqrt{q}}={1 \over x}
\end{equation*}
The equation \eqref{31} is an exact analytic constraint, allowing the location 
of 
the transition point $K_t$ (given by $x=1$) on the basis of the non-analyticity 
of $\phi(K)$ at this point \cite{Baxter82}. However, besides this formal 
aspect, 
we have no trace, in the above relation, of the mechanism leading to a 
transition from the disordered to the ordered phase.
\par Now, we will see that further insights and additional aspects of duality 
are provided by the variational method, by means of the ansatz's introduced in 
the previous section.
\par First of all we remark that, due to the fact that the ordered and 
disordered regions are described by a unique vector $\psi_1(\sigma)$ for finite 
$m$, the two {\it charts} of a proper ansatz should share some common property, 
as an expression of their common origin. As a matter of fact there is a 
property 
that allows to associate univocally $\tilde\psi_o^{(i)}(\sigma)$ with $L^{1 
\over 2}\tilde\psi_d^{(i)}(\sigma)=\tilde\psi_d^{(i+1)}(\sigma)$, leading us to 
consider, as proper ansatz's with two {\it charts}, the sequence
\begin{equation}
\tilde\psi^{(1)}(\sigma)=(\tilde\psi_d^{(2)}(\sigma),\tilde\psi_o^{(1)}(\sigma)
)
,\quad \dots \quad,
\tilde\psi^{(i)}(\sigma)=(\tilde\psi_d^{(i+1)}(\sigma), 
\tilde\psi_o^{(i)}(\sigma)),
\label{32}
\end{equation}
and then the sequence of estimates $\phi^{(i)}(K)$ for $\phi(K)$, given by
\begin{equation}
\phi^{(i)}(K) = \max (\phi_d^{(i+1)}(K),\phi_o^{(i)}(K)) \qquad (i=1,2,\dots)
\label{33}
\end{equation}
This property, shared by $\tilde\psi_o^{(i)}(\sigma)$ and  
$\tilde\psi_d^{(i+1)}(\sigma)$, is an algebraic structure, as it will be shown 
in the following.
\par We first consider $\tilde\psi_d^{(\nu)}(\sigma)$, for $\nu \geq 2$. We have 
seen that this vector defines the Potts$(\nu;A)$ model on a strip 
$\Sigma_{\nu}$. Let us call $\tau=(t_1,t_2,\dots,t_\nu)$ a spin configuration 
on 
a row of $\Sigma_{\nu}$. The transfer matrix $\ell_{\nu} (A,K)$, which connects 
two adjacent rows, can be written in the form \cite{Baxter82}
\begin{equation}
\ell_{\nu} (A,K) = V_\nu W_\nu \equiv q^{\nu \over 2} a^2 x^{\nu - 2} \tilde 
\ell_{\nu} (A,K),
\label{34}
\end{equation}
with
\begin{align}
V_\nu &= (I + x U_\nu^{(2)})(I + x U_\nu^{(4)}) \dots (I + x U_\nu^{(2\nu - 
2)}) 
\qquad (\nu \geq 2) \notag \\
W_\nu &=
  \begin{cases}
    q a^2 (I + {1 \over a} U_2^{(1)})(I + {1 \over a} U_2^{(3)}) & (\nu=2) \\
    q^{\nu \over 2} a^2 x^{\nu - 2}(I + {1 \over a} U_\nu^{(1)})(I + {1 \over 
a}  
    U_\nu^{(2\nu-1)})(I + {1 \over x} U_\nu^{(3)}) \dots (I + {1 \over x}
    U_\nu^{(2\nu-3)}) & (\nu \geq 3),
  \end{cases} 
\label{35}
\end{align}
where
\begin{equation*}
a={{e^{{A+K} \over 2}-1} \over \sqrt{q}}
\end{equation*}
and $U_\nu^{(1)},U_\nu^{(2)},\dots,U_\nu^{(2\nu-1)}$ are $q^\nu \times q^\nu$ 
matrices defined by
\begin{equation*}
  \begin{split}
   &U_\nu^{(2i-1)}(\tau,\tau') = {1 \over \sqrt{q}} 
   \prod \begin{Sb} j=1 \\ j \neq i \end{Sb} \delta_{t_j,t'_j} \qquad 
   (i=1,2,\dots,\nu) \\
   &U_\nu^{(2i)}(\tau,\tau') = \sqrt{q} \delta_{t_i,t_{i+1}}
   \prod _{j=1}^\nu \delta_{t_j,t'_j} \qquad (j=1,2,\dots,\nu-1)
   \end{split}
\end{equation*}
They satisfy the relations
\begin{equation}
  \begin{split}
    &(U_\nu^{(i)})^2 = \sqrt{q} U_\nu^{(i)} \quad (i=1,2,\dots,2\nu-1) ;\\
    &U_\nu^{(i)} U_\nu^{(j)} = U_\nu^{(j)} U_\nu^{(i)} \quad
    (|i-j| \geq 2);\\
    &U_\nu^{(i)} U_\nu^{(i+1)} U_\nu^{(i)} = U_\nu^{(i)} \quad
     (i=1,2,\dots,2\nu-2);\\
    &U_\nu^{(i)}U_\nu^{(i-1)} U_\nu^{(i)} = U_\nu^{(i)} \quad
     (i=1,2,\dots,2\nu-1)
   \end{split}
\label{36}
\end{equation}
\par These relations define the Temperley-Lieb algebra generated by 
$U_\nu^{(1)},U_\nu^{(2)}, \dots, U_\nu^{(2\nu-1)}$, and then all the 
eigenvalues of $\ell_\nu(A,K)$, with exception of their degeneracies.
\par Now we come to $\tilde\psi_o^{(\nu)}(\sigma) (\nu=1,2,\dots)$, which 
defines the Potts$(\nu+2;+\infty,H)$ model. It is useful to consider this model 
as the limit, for $R \to +\infty$, of Potts$(\nu+2;R,H)$, obtained through the 
substitution of $\tilde\psi_o^{(1)}(\sigma)$ with 
$\tilde\psi_o^{(1)}(\sigma;R)$ 
of the equation \eqref{26}. Then, following the previous procedure, we have 
that 
to Potts$(\nu;R,H) (\nu \geq 3)$ is associated a $q^\nu \times q^\nu$ transfer 
matrix $\ell_\nu(R,H,K)$, which can be written in the form
\begin{equation}
\ell_\nu(R,H,K)=\bar V_\nu \bar W_\nu(R),
\label{37}
\end{equation}
with
\begin{align}
 \bar V_\nu &= 
  \begin{cases}
    (I + h U_\nu^{(2)})(I + h U_\nu^{(2\nu-2)}) (I + x U_\nu^{(4)}) 
    \dots (I + x U_\nu^{(2\nu - 4)}) \qquad &(\nu \geq 4) \\
    (I + h U_3^{(2)})(I + h U_3^{(4)}) \qquad &(\nu =3)
  \end{cases} \notag \\
 \bar W_\nu &= e^{-2R} q^{\nu \over 2} \rho ^2 x^{\nu - 2}(I + {1 \over \rho} 
 U_\nu^{(1)})(I + {1 \over \rho}U_\nu^{(2\nu-1)})(I + {1 \over x} U_\nu^{(3)})  
 \dots (I + {1 \over x}U_\nu^{(2\nu-3)}) \quad (\nu \geq 3),
\label{38}
\end{align}
where
\begin{equation*}
h={{e^H-1} \over \sqrt{q}} \quad , \quad \rho={{e^R-1} \over \sqrt{q}}
\end{equation*}
Taking the limit $R \to +\infty$, we obtain
\begin{equation}
\ell_\nu(+\infty,H,K)=\bar V_\nu \bar W_\nu(+\infty) \equiv q^{{\nu-2}\over 2} 
x^{\nu-2} \tilde\ell_\nu(+\infty,H,K)
\label{39}
\end{equation}
with
\begin{equation}
\bar W_\nu(+\infty) = q^{{\nu-2}\over 2} x^{\nu-2} (I + {1 \over x} 
U_\nu^{(3)}) 
\dots (I + {1 \over x}U_\nu^{(2\nu-3)})
\label{40}
\end{equation}
So, the matrices $U_\nu^{(1)}$ and $U_\nu^{(2\nu-1)}$ being suppressed by the 
limiting procedure, we remain only with the $2(\nu-1)-1 \quad q^\nu \times 
q^\nu$ matrices $U_\nu^{(2)},U_\nu^{(3)},\dots,U_\nu^{(2\nu-3)},U_\nu^{(2\nu-2)}$. However, 
if we define
\begin{equation*}
\bar U_\nu^{(i)}=U_\nu^{(i+1)} \quad , \quad (i=1,2,\dots,2(\nu-1)-1)
\end{equation*}
we see from \eqref{36}, that these $2(\nu-1)-1 \quad q^\nu \times q^\nu$ 
matrices $\bar 
U_\nu^{(i)}$ satisfy the same algebraic relations which are satisfied by the 
$q^{\nu-1} \times q^{\nu-1}$ matrices $U_{\nu-1}^{(i)} 
(i=1,2,\dots,2(\nu-1)-1)$. 
We only have with the $\bar U_\nu^{(i)}$ a representation of different 
dimensionality of the same algebraic structure. Together with the structure of 
$\tilde\ell_\nu(+\infty,H,K)$ and $\tilde\ell_{\nu-1}(A,K)$, this is the common 
property which allows to couple the vector 
$\tilde\psi_o^{(\nu-2)}(\sigma) (\nu\geq 3)$ with the vector 
$\tilde\psi_d^{(\nu-1)}(\sigma)$, leading then to the sequence \eqref{32}.
\par As a matter of fact, it follows from \eqref{34}, \eqref{35}, \eqref{38}, 
\eqref{39} and \eqref{40} that the eigenvalues of $\tilde\ell_\nu(+\infty,H,K)$ 
coincide with those of 
$\tilde\ell_{\nu-1}(A_D(H,K^*),K^*)$, where $A_D(H,K^*)$ is defined by
\begin{equation}
{{e^{{A_D(H,K^*) + K^*}\over 2} - 1} \over \sqrt{q}} = {1 \over h}
\label{41}
\end{equation}
In particular, we have
\begin{equation}
\lambda_1^{(\nu)}(+\infty,H,K) = {h^2 \over \sqrt{q}} x^{2\nu-5} 
\lambda_1^{(\nu-1)}(A_D(H,K^*),K^*) \qquad (\nu\geq 3),
\label{42}
\end{equation}
which is a duality relation for the partition function of 
Potts$(\nu;+\infty,H)$ and Potts$(\nu-1;A)$.
\par As a consequence, coming back to the variational method, we obtain the 
duality relation for the R.R. quotients
\begin{equation}
{\lambda_1^{(i+3)}(+\infty,H,K) \over \lambda_1^{(i+2)}(+\infty,H,K)} = x^2 
{\lambda_1^{(i+2)}(A_D(H,K^*),K^*) \over \lambda_1^{(i+1)}(A_D(H,K^*),K^*)}
\label{43}
\end{equation}
Since, as $H$ goes from zero to $+\infty$, $A_D(H,K^*)$ goes from $+\infty$ to 
$-K^*$, we deduce from \eqref{43}, \eqref{19}, \eqref{28} that
\begin{equation*}
\phi_o^{(i)}(K) = \log x^2 + \phi_d^{(i+1)}(K^*),
\end{equation*}
or, equivalently,
\begin{equation}
\phi_d^{(i+1)}(K) = \log x^2 + \phi_o^{(i)}(K^*),
\label{44}
\end{equation}
the relative $\sup$ being attained at {\it dual} points, according to 
\eqref{41}. 
So $\tilde\psi_o^{(i)}(\sigma)$ and $\tilde\psi_d^{(i+1)}(\sigma)$ are exactly 
compatible.
\par The equation \eqref{44} is the variational version of \eqref{31}. Indeed, 
from 
\eqref{44} it follows that
\begin{equation}
\phi_o^{(i)}(K)-\phi_d^{(i+1)}(K) = \phi_d^{(i+1)}(K^*)-\phi_o^{(i)}(K^*)
\label{45}
\end{equation}
Therefore, if at a point $K$ we have, for example, $\phi_d^{(i+1)}(K) > 
\phi_o^{(i)}(K)$, then at the dual point it results that
$\phi_o^{(i)}(K^*)>\phi_d^{(i+1)}(K^*)$. 
So we conclude that the estimates $\phi^{(i)}(K)$ given by \eqref{33}, satisfy 
exactly the duality relation \eqref{31}, that is
\begin{equation}
\phi^{(i)}(K) = \log x^2 + \phi^{(i)}(K^*) \qquad (i=1,2,\dots).
\label{46}
\end{equation}
We see that the mechanism of non-analyticity at $K_t$ can be described as a 
crossing phenomenon, involving $\phi_o^{(i)}(K)$ and $\phi_d^{(i+1)}(K)$.

\section{Two-parameters ansatz's}
The above results can be extended to two-parameters ansatz, which are useful to 
consider both from the computational and variational point of view.
\par To fix our ideas, let us suppose that we need to go beyond the estimates 
$\phi_d^{(2)}(K)$ and $\phi_d^{(3)}(K)$. We can consider then 
$\phi_d^{(4)}(K)$. The relative calculations require strips of four and five 
columns, to be compared with the case of three and four columns needed for 
$\phi_d^{(3)}(K)$. However we can also improve $\phi_d^{(3)}(K)$ without 
changing the number of columns  and the size of the involved transfer matrices, 
but by adding a further parameter. In fact, by making use of the procedure of 
an approximate description of the averaging process, we can introduce the 
two-parameters ansatz for $\psi_1(\sigma)$
\begin{equation}
\tilde\psi_d^{'(3)}(\sigma)=\Bigl( \sum_{\sigma'} e^{A \sum_{i=1}^{m-1} 
\delta_{s'_i,s'_{i+1}} + B \sum_{i=1}^m \delta_{s'_i,s_i}} \Bigr) e^{{K \over 
2} {\sum_{i=1}^{m-1} \delta_{s_i,s_{i+1}}}}
\label{47}
\end{equation}
with $A,B \geq 0$.
\begin{figure}[ht]
\begin{center}
\mbox{\epsfig{file=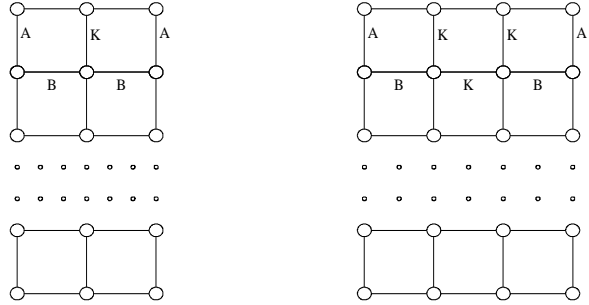,width=13cm,height=7cm}}
\end{center}
\caption{The $3$ and $4$ columns Potts models originated by the two-parameters ansatz \eqref{47}. They correspond to the partition functions $Z_d^{(3)}$ and $Z_d^{(4)}$ leading to our estimate $\phi_d^{'(3)}$ of the free energy in the disordered phase.}
\label{fig1}
\end{figure}
\par
\begin{figure}[ht]
\begin{center}
\mbox{\epsfig{file=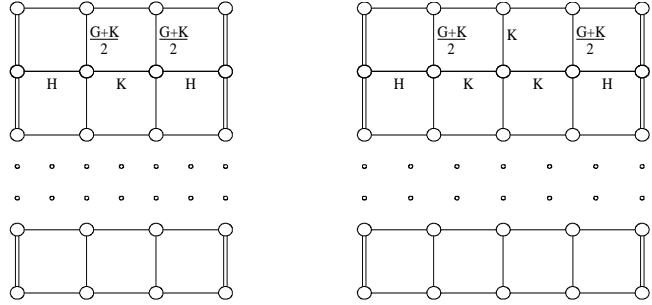,width=13cm,height=7cm}}
\end{center}
\caption{The $4$ and $5$ columns Potts models originated by the two-parameters 
ansatz \eqref{50}. They correspond to the partition functions $Z_o^{(4)}$ and 
$Z_o^{(5)}$ leading to our estimate $\phi_o^{'(2)}$ of the free energy in the 
ordered phase. Double lines connect spins constrained to be in the same state.}
\label{fig2}
\end{figure}

 Then, by considering $(\tilde \psi_d^{'(3)}, L \tilde \psi_d^{'(3)}) / 
(\tilde \psi_d^{'(3)}, \tilde \psi_d^{'(3)})$, we are led to the estimate of 
$\phi (K)$
\begin{equation}
\phi_d^{'(3)}(K) = \sup_{A,B} \lim_{m \to \infty} {1 \over m} \log 
{Z_d^{(4)}(A,B,K) \over Z_d^{(3)}(A,B,K)},
\label{48}
\end{equation}
where $Z_d^{(i)}(A,B,K) (i=3,4)$ are the partition functions of the Potts 
models shown in Fig. (\ref{fig1}). Therefore, we deduce from \eqref{19} and 
\eqref{48}
\begin{equation}
\phi_d^{(3)}(K) \leq \phi_d^{'(3)}(K) \leq \phi(K)
\label{49}
\end{equation}
We can improve analogously $\phi_o^{(2)}(K)$, which is determined through 
$\tilde \psi_o^{(2)}(\sigma) = L^{1 \over 2} \tilde \psi_o^{(1)}(\sigma)$, by 
introducing the two-parameter ansatz
\begin{equation}
\tilde \psi_o^{'(2)}(\sigma) = L^{1 \over 2} \tilde \psi_o^{'(1)}(\sigma)
\label{50}
\end{equation}
with
\begin{equation}
\tilde \psi_o^{'(1)}(\sigma) = \Bigl( \sum_{\sigma'} \Pi(\sigma',\sigma) \Bigr) 
e^{{G \over 2} \sum_{i=1}^{m-1} \delta_{s_i,s_{i+1}}} \qquad (G \geq 0),
\label{51}
\end{equation}
where $\Pi(\sigma',\sigma)$ is given by \eqref{21}.
Through $(\tilde \psi_o^{'(2)}, L \tilde \psi_o^{'(2)}) / (\tilde 
\psi_o^{'(2)}, \tilde \psi_o^{'(2)})$, we obtain the estimate
\begin{equation}
\phi_o^{'(2)}(K) = \sup_{G,H} \lim_{m \to \infty} {1 \over m} \log 
{Z_o^{(5)}(G,H,K) \over Z_o^{(4)}(G,H,K)},
\label{52}
\end{equation}
where $Z_o^{(i)}(G,H,K) (i=4,5)$ are the partition functions of the Potts 
models shown in Fig. (\ref{fig2}).
\par By making use again of the generators of the Temperley-Lieb algebra we 
have, as in \eqref{44},
\begin{equation}
\phi_d^{'(3)}(K) = \log x^2 + \phi_o^{'(2)}(K^*),
\label{53}
\end{equation}
so that also the estimate of $\phi(K)$,
\begin{equation}
\phi^{'(2)}(K) = \max (\phi_d^{'(3)}(K),\phi_o^{'(2)}(K)),
\label{54}
\end{equation}
satisfies the duality relation \eqref{31}. As a matter of fact, the parameters 
$H$ and $G$ are the {\it dual} expression of $A$ and $B$, respectively.
\par In principle, further improved two-parameters ansatz's can be obtained 
from the action of$(L^{1 \over 2})^i$ on the vector $\tilde \psi^{'(2)} 
(\sigma) = (\tilde \psi_d^{'(3)} (\sigma),\tilde \psi_o^{'(2)} (\sigma))$, 
leading to the sequence of estimates $\phi^{'(i+2)}(K)$.

\section{Some numerical results and comments}
Now we discuss the effective aspects of the variational method presented in the 
previous sections, by considering the estimates of lower order.
\par We have calculated explicitly 
$\phi_d^{(2)}(K),\phi_d^{(3)}(K),\phi_d^{'(3)}(K),\phi_o^{(1)}(K), 
\phi_o^{(2)}(K)$ and $\phi_o^{'(2)}(K)$, obtaining then $\phi^{(1)}(K), 
\phi^{(2)}(K)$ and $\phi^{'(2)}(K)$, according to \eqref{33} and \eqref{54}. We 
have also considered $\phi_d^{(1)}(A,K)$, which does not have a {\it dual} 
partner in our scheme, but, in any case, allows to obtain the simplest 
approximation to $\phi(K)$, for every $K$. As a matter of fact, 
$\phi_d^{(1)}(A,K)$ has a finite absolute maximum point $\bar A(K)$ for $K < 
\bar K_t$, while for $K > \bar K_t$ the absolute maximum is taken at $A \to 
+\infty$. These points are also relative maximum points in a neighbourhood of 
$\bar K_t$, at its right and at its left, respectively. $\bar K_t$ is an 
estimate of the transition point $K_t$; at $\bar K_t$ we have a crossing 
between $\phi_d^{(1)}(\bar A,K)$ and $\phi_d^{(1)}(+\infty,K)$.
\par The simplest estimate of our variational method is then given by
\begin{equation}
\phi^{(0)}(K) = \max \Bigl( \phi_d^{(1)}(\bar A(K),K),\phi_d^{(1)}(+\infty,K) 
\Bigr).
\label{55}
\end{equation}
\begin{figure}[ht]
\begin{center}
\mbox{\epsfig{file=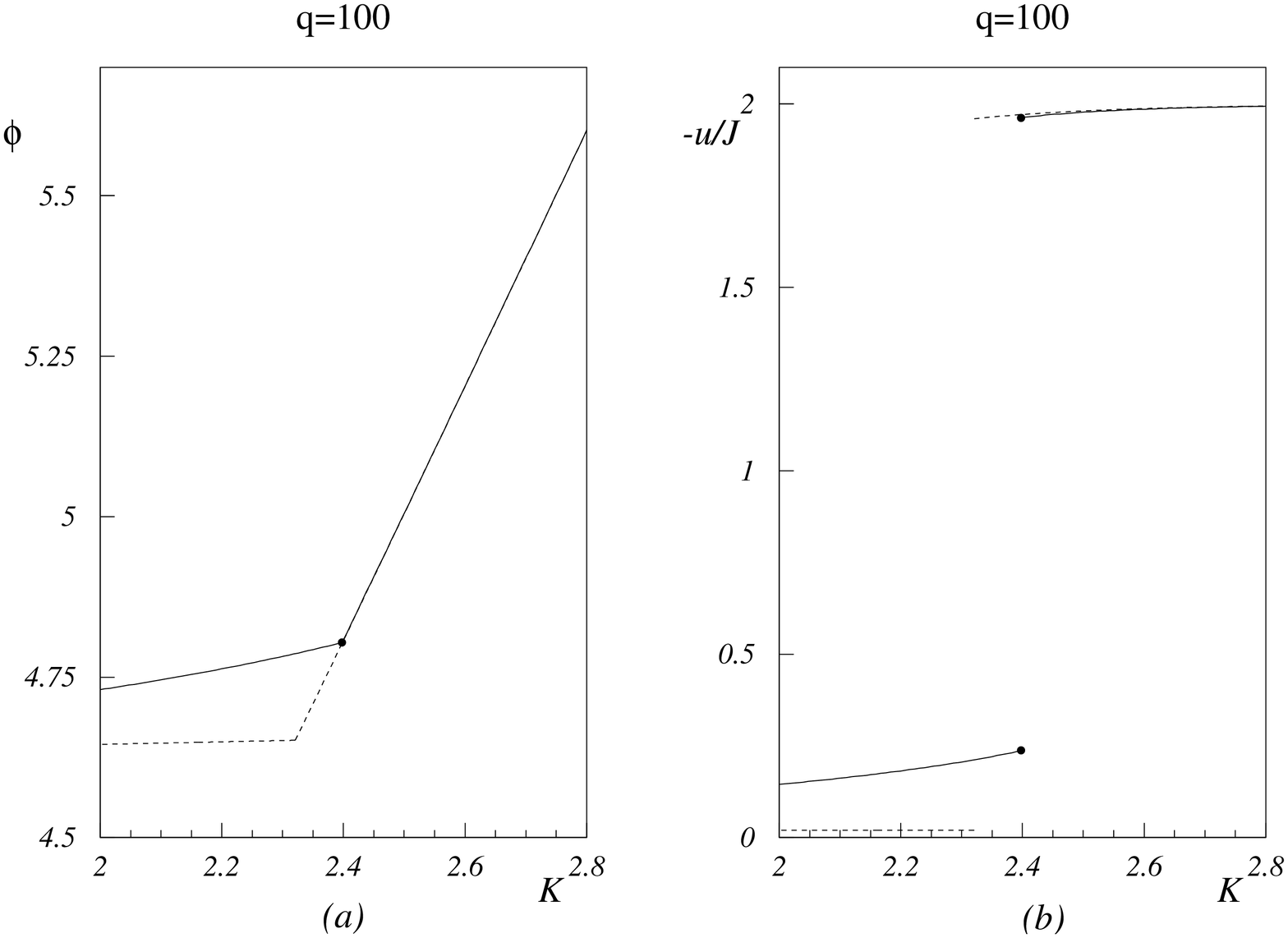,width=16cm,height=10cm}}
\end{center}
\caption{Our simplest estimate (full lines) of the free energy (a) and of the internal energy (b) at $q=100$ are compared with the results from the Mean Field Theory (dashed lines). Also shown are the exact values at the transition point (small circles).}
\label{fig3}
\end{figure}
It turns out that $\phi^{(0)}(K)$ satisfies quite well, even if approximately, 
the duality relation \eqref{31}, when $q$ is large. As a matter of fact, by 
making a comparison with the known exact results, we have that $\phi^{(0)}(K)$ 
provides a quite reliable description, for every $K$, of the properties of the 
2D square Potts model for $q \geq 50$, the description being more and more 
accurate as $q$ increases. In this respect, our simplest approximation behaves 
like Mean Field Theory. However, from a quantitative point of view, we get a 
net improvement, since it turns out that the Mean Field estimates, which are 
exact in the limit $q \to +\infty$, differ significantly from the known exact 
results already at $q$ of order $10^3$. Indeed, in the large $q$ expansions, we 
have power series in $1/\sqrt{q}$ with very large coefficients 
\cite{BLM96}, which are responsible of such deviations. In Figures 
(\ref{fig3})a and (\ref{fig3})b we plot, for $q=100$, $\phi^{(0)}(K)$ and the 
relative internal energy $u^{(0)}(K)$, given by
\begin{equation*}
{u^{(0)}(K) \over {-J}} = {d \over dK} \phi^{(0)}(K),
\end{equation*}
and we make a comparison with the analogous Mean Field quantities 
$\phi_{MF}(K)$ and $u_{MF}(K)$; the small circles are the known exact values.
\par From a computational point of view, in order to avoid $q^\nu \times q^\nu$ 
matrices of very high order, especially if $q$ is large, we have made use of 
the relationship between the reduced transfer matrices and the matrices 
$U_\nu^{(1)}, \dots, U_\nu^{(2\nu-1)}$ which satisfy the relations \eqref{36} 
(see \eqref{34},  \eqref{35}, \eqref{37} and \eqref{38}). Now the relations 
\eqref{36} are also satisfied by the $4^\nu \times 4^\nu$ matrices 
$U_\nu^{(i)}$ given by \cite{Baxter82}
\begin{equation}
U_\nu^{(i)}(\alpha_1,\alpha_2, \dots, \alpha_{2\nu}|\alpha'_1,\alpha'_2, \dots, 
\alpha'_{2\nu}) = \delta_{\alpha_1 \alpha'_1} \dots \delta_{\alpha_{i-1} 
\alpha'_{i-1}} h_{\alpha_i \alpha_{i+1}} h_{\alpha'_i \alpha'_{i+1}} 
\delta_{\alpha_{i+2} \alpha'_{i+2}} \dots \delta_{\alpha_{2\nu} \alpha'_{2\nu}} 
\label{56}
\end{equation}
$(i=1,2,\dots,2\nu-1)$, where $\alpha_i=\pm 1, \alpha'_i=\pm 1$,
\begin{equation*}
h_{+ +} = h_{- -} = 0, \qquad h_{+ -} = e^{-{\lambda \over 2}}, \qquad h_{- +} 
= e^{\lambda \over 2}
\end{equation*}
with $2 \cosh \lambda = q^{1/2}$.
This is the 6-vertex representation of the Temperley-Lieb algebra.
\par The eigenvalues of our reduced transfer matrices have been determined by 
making use of the representation \eqref{56}, which allows to translate a Potts 
strip $\Sigma_i$ into an ice-type (or 6-vertex) strip $\Sigma'_i$ and, as a 
by-product, to consider also non integer values of $q$.
\par As we had to expect, the results of our calculations show that $K_t$ is 
the only crossing point of $\phi_d^{(i+1)}(K)$ with $\phi_o^{(i)}(K) (i=1,2)$  
and also of $\phi_d^{'(3)}(K)$ with $\phi_o^{'(2)}(K)$. Furthermore  
$\phi_d^{(i+1)}(K) > \phi_o^{(i)}(K), \quad \phi_d^{'(3)}(K) > 
\phi_o^{'(2)}(K)$ for $K<K_t$, so that
\begin{equation}
\phi^{(i)}(K) = 
  \begin{cases}
    \phi_d^{(i+1)}(K)& K \leq K_t \\
    \phi_o^{(i)}(K)& K \geq K_t
  \end{cases}
  \begin{gathered}
   ,\qquad \phi^{'(2)}(K)=
     \begin{cases}
       \phi_d^{'(3)}(K)& K \leq K_t \\
       \phi_o^{'(2)}(K)& K \geq K_t
     \end{cases}
  \end{gathered}
\label{57}
\end{equation}
\begin{table}
\caption{ Our predictions for the free energy $\phi$, the internal energy in the disorder phase $u_d$ and the specific heat in the ordered phase $C_o$ calculated at the transition point $K_t$. The exact values for the free internal energy and the internal energy are also reported.}

\begin{tabular}{|c|l|l|l|l|l|l|}
\hline 
q & 5 & 10 & 15 & 20 & 30 & 100 \\
\hline 
$\phi^{(0)}(K_t)$   & 2.403806 & 2.895964 & 3.202524 & 3.428507 & 3.758386 & 4.803843 \\
$\phi^{(1)}(K_t)$   & 2.407461 & 2.898352 & 3.204045 & 3.429543 & 3.758951 & 4.803915 \\
$\phi^{(2)}(K_t)$   & 2.408681 & 2.899057 & 3.204432 & 3.429775 & 3.759055 & 4.803922 \\
$\phi^{'(2)}(K_t)$  & 2.409378 & 2.899363 & 3.204563 & 3.429841 & 3.759078 & 4.803923 \\
$\phi^{''(2)}(K_t)$ & 2.410306 & 2.899597 & 3.204631 & 3.429867 & 3.759084 & 4.803923 \\
$\phi(K_t)$         & 2.409849 & 2.899522 & 3.204615 & 3.429862 & 3.759083 & 4.803923 \\
\hline
$-u_d^{(0)}(K_t)/J$   & 1.189595 & 0.868464 & 0.701564 & 0.597935 & 0.474256 & 0.236934 \\
$-u_d^{(1)}(K_t)/J$   & 1.249544 & 0.911736 & 0.728555 & 0.615741 & 0.483472 & 0.237961 \\
$-u_d^{(2)}(K_t)/J$   & 1.282729 & 0.933513 & 0.739933 & 0.622203 & 0.486151 & 0.238115 \\
$-u_d^{'(2)}(K_t)/J$  & 1.317667 & 0.949882 & 0.746210 & 0.625067 & 0.487035 & 0.238139 \\
$-u_d^{''(2)}(K_t)/J$ & 1.413784 & 0.971819 & 0.751328 & 0.626773 & 0.487391 & 0.238142 \\
$-u_d(K_t)/J$         & 1.420754 & 0.968203 & 0.750492 & 0.626529 & 0.487353 & 0.238142\\
\hline
$C^{(1)}_o(K_t)$   &  4.5049 &  5.4063 & 4.6024 & 3.8641 & 2.9059 & 1.1762 \\
$C^{(2)}_o(K_t)$   &  6.2350 &  7.2927 & 5.6956 & 4.5144 & 3.1923 & 1.1969 \\
$C^{'(2)}_o(K_t)$  &  9.9421 &  9.8526 & 6.6928 & 4.9700 & 3.3350 & 1.2013 \\
$C^{''(2)}_o(K_t)$ & 30.6324 & 15.7009 & 7.9572 & 5.3728 & 3.4165 & 1.2022 \\
\hline
\end{tabular}
\label{table1}
\end{table}  
In order to see the degree of accuracy which we obtain, we compare in 
Table \ref{table1} our results at $K_t$, for several $q$, with the known exact 
values $\phi(K_t)$ and $-u_d(K_t)/J$. Of course, due to the 
duality relation, we reproduce exactly, through $\phi^{(1)}(K),\phi^{(2)}(K)$ 
and $\phi^{'(2)}(K)$, the mean value $(u_d(K_t)+u_o(K_t))/2$.
\par We see that, for every $q>4$, $\phi^{(0)}(K_t), \phi^{(1)}(K_t), 
\phi^{(2)}(K_t)$ and $\phi^{'(2)}(K_t)$ provide a monotonous sequence of very 
accurate estimates  of $\phi(K_t)$, the accuracy being better as $q$ increases. 
At $q=5$, the errors of these estimates are of $0.25\%, 0.10\%, 0.05\%,0.02\%$ 
respectively, while at $q=10$, the sequence of the errors is 
$0.12\%,0.04\%,0.02\%,0.005\%$. As a matter of fact, by analysing the stability 
of our results for $K \not= K_t$, we obtain that these errors are still rapidly 
decreasing, as we move from $K_t$.
\par For the internal energy, our predictions are characterized by larger 
errors, provided $q$ is low and $K$ is very near to $K_t$. But our last 
estimate, through the two-parameters variational ansatz, is quite accurate for 
every $K$ and every $q\gtrsim 15$. At $q=20$, $-u_d^{'(2)}(K_t)/J$ gives the 
exact value with an error of $0.1\%$. On the other hand, we obtain the errors of 
$7\%$ and $2\%$ at $q=5$ and $q=10$, respectively. 
\begin{figure}[ht]
\begin{center}
\mbox{\epsfig{file=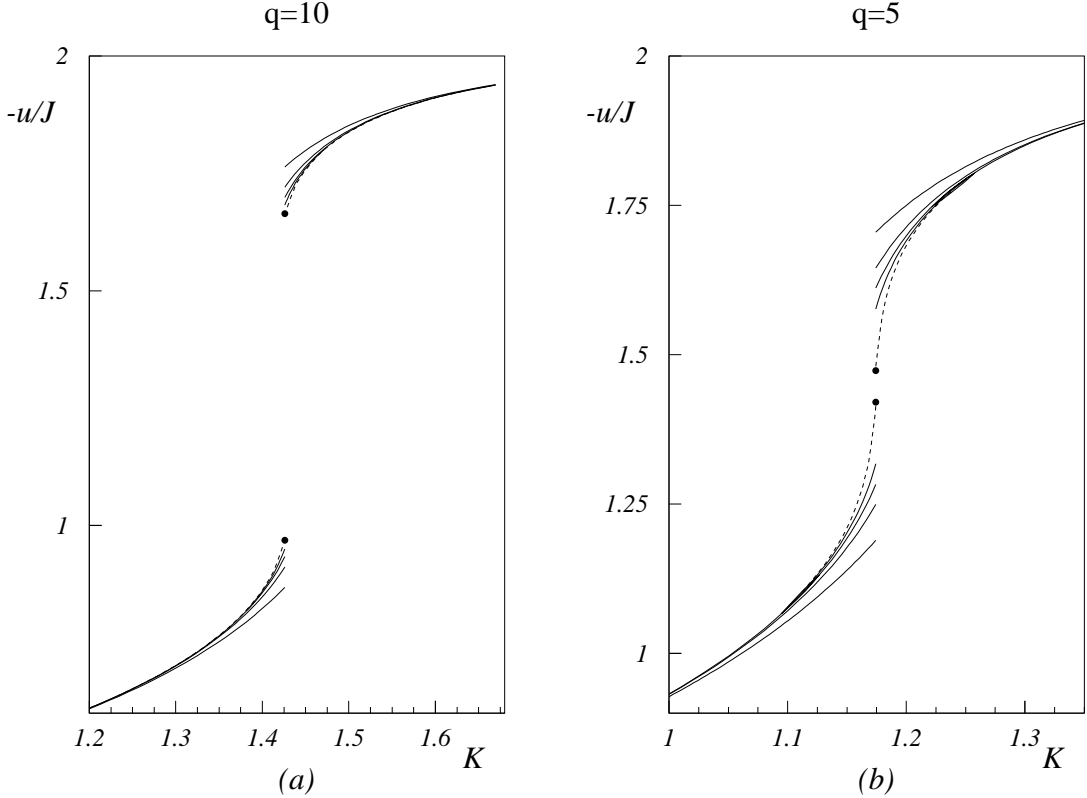,width=16.4cm}}
\end{center}
\caption{The monotonous sequence of our estimates $-u^{(1)}(K)/J, -u^{(2)}(K)/J$ and $-u^{'(2)}(K)/J$ of the internal energy at $q=10$ (a) and $q=5$ (b). The dashed line represents $-u^{''(2)}(K)/J$, obtained from  eq.\eqref{58}. Also shown are the exact values at the transition point (small circles).}
\label{fig4}
\end{figure}
\begin{figure}[ht]
\begin{center}
\mbox{\epsfig{file=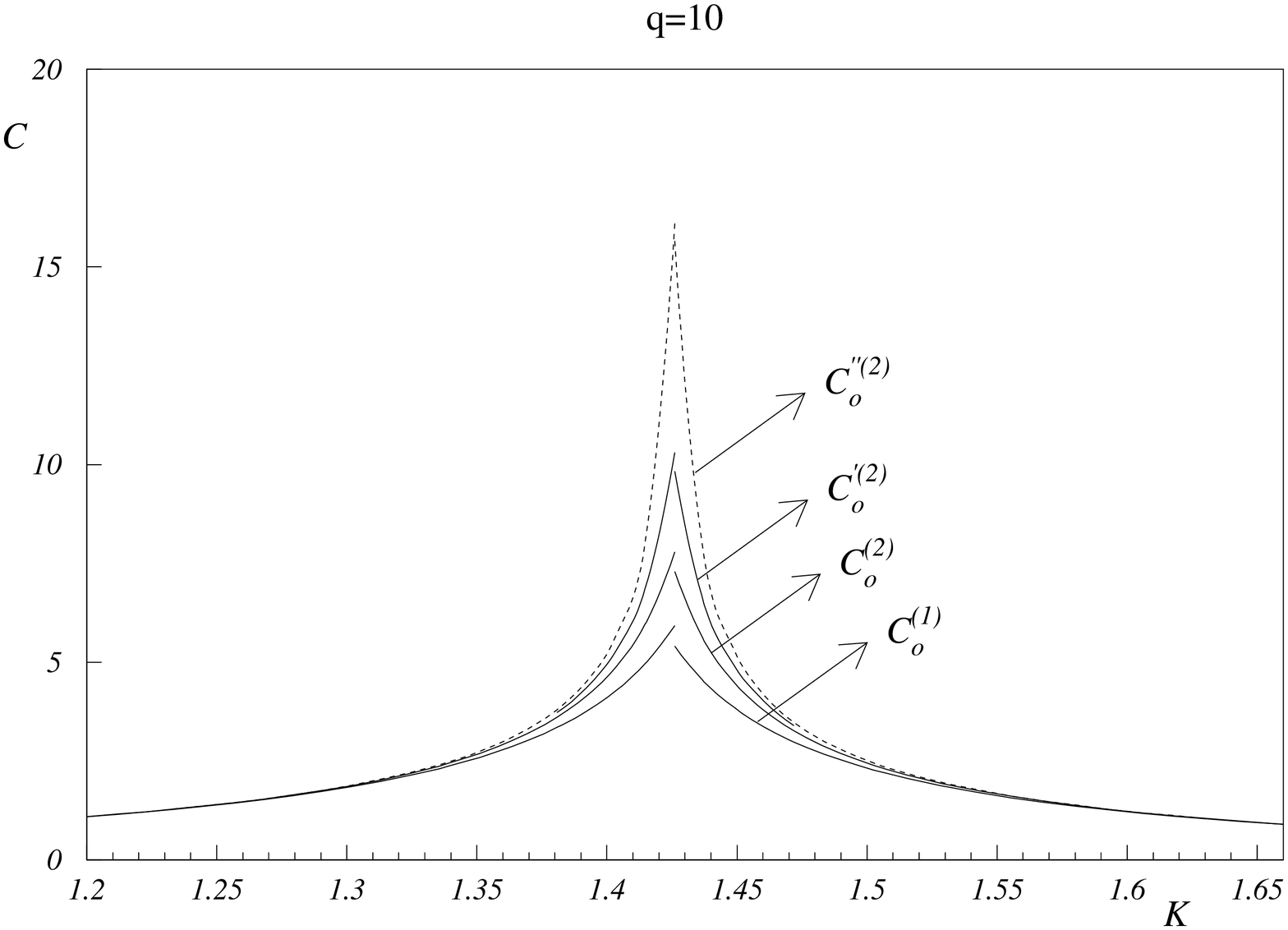,width=16cm,height=10cm}}
\end{center}
\caption{The monotonous sequence of our estimates $C^{(1)}(K), C^{(2)}(K)$ and $C^{'(2)}(K)$ of the specific heat at $q=10$. The dashed line represents $C^{''(2)}(K)$, obtained from eq.\eqref{58}.}
\label{fig5}
\end{figure}
However, as we move from $K_t$, our estimates 
of the internal energy are again really quite accurate, even for low $q$'s. 
This is shown in Fig.(\ref{fig4})a and Fig.(\ref{fig4})b, where we plot the 
functions $-u^{(1)}(K)/J, -u^{(2)}(K)/J$ and $-u^{'(2)}(K)/J$.
\par Through  $\phi^{(i)}(K)\quad (i=0,1,2)$ and $\phi^{'(2)}(K)$ we have also 
determined the estimates $C^{(i)}(K)$ and $C^{'(2)}(K)$ of the specific heat 
$C(K)$, which is not exactly known, even at $K_t$. The sequence  $C^{(0)}(K), 
C^{(1)}(K), C^{(2)}(K), C^{'(2)}(K)$ is monotonous and rapidly converging for
every $K$, provided $q\gtrsim 30$. For these $q$'s, $C^{'(2)}(K)$ provides a 
quite reliable estimate of $C(K)$ for every $K$. This is also true for 
$C^{(0)}(K), C^{(1)}(K)$ and $C^{(2)}(K)$ as soon as $q$ increases. For lower 
values of $q$, we have to move from $K_t$ in order to get the same accuracy. In 
Fig.(\ref{fig5}) we plot $C^{(1)}(K), C^{(2)}(K)$ and $C^{'(2)}(K)$ at $q=10$, 
while in Table \ref{table1} we give, for several $q$, $C^{(1)}_o(K_t), 
C^{(2)}_o(K_t)$ and $C^{'(2)}_o(K_t)$, that is the previous estimates at $K_t$, 
evaluated in the ordered phase (the predictions for the disordered phase can be 
obtained from these through duality).
\par On the other hand, we obtain a considerable improvement in our estimates of 
the internal energy and of the specific heat for lower $q$ and $K$ very near to 
$K_t$, by accelerating the convergence of the already rapidly convergent 
sequence of our estimates of the free energy. Starting from $\phi^{(1)}(K)$, 
$\phi^{(2)}(K)$ and $\phi^{'(2)}(K)$ and by making use of the Aitken $\Delta^2$ 
method (or of the $[1,1]$ Pad\`e approximant), we deduce the new extimate
\begin{equation}
\phi^{''(2)}(K)=\phi^{(1)}(K)+{{\phi^{(2)}(K)-\phi^{(1)}(K)} 
\over {1-{{\phi^{'(2)}(K)-\phi^{(2)}(K)} \over {\phi^{(2)}(K)-\phi^{(1)}(K)}}}}.
\label{58}
\end{equation}
We call $u^{''(2)}(K)$ and $C^{''(2)}(K)$ the internal energy and the specific 
heat calculated through $\phi^{''(2)}(K)$. The improvement  which we obtain can 
be checked by comparing $u_d^{''(2)}(K_t)$ with the known exact value. We give 
in Table \ref{table1} $\phi^{''(2)}(K_t)$, $u_d^{''(2)}(K_t)$ and 
$C_o^{''(2)}(K_t)$ for several $q$. We see that $-u_d^{''(2)}(K_t)/J$ reproduces 
$-u_d(K_t)/J$ within the errors of $0.5\%$ and $0.4\%$ at $q=5$ and $q=10$, 
respectively, a net improvement with respect $-u_d^{'(2)}(K_t)/J$. On the other 
hand, when $u_d^{'(2)}(K)$ and $C^{'(2)}(K)$ are already quite reliable, 
$\phi^{''(2)}(K)$ does not lead to significant corrections. For $q=10$ and 
$q=5$, the function $-u_d^{''(2)}(K)/J$ is plotted in Fig.(\ref{fig4})a and 
Fig.(\ref{fig4})b, where its role is made quite evident. A stronger correction 
is obtained for specific heat, as shown in Fig.(\ref{fig5}), where we plot 
$C^{''(2)}(K)$, for $q=10$. Our estimate of $16$ for $C_o(K_t)$ at $q=10$, 
agrees essentially with the estimate of $18$ obtained from large-$q$ expansions 
\cite{BLM96}, which is contradicted by the value $32$ deduced from low 
temperature series \cite{BEG94}. For $q=15,20,30$ the results of ref. 
\cite{BLM96} are confirmed quite well by our $C_o^{''(2)}(K_t)$.
\par So, we see that the variational approach presented in this paper, besides 
providing exact predictions and physical or mathematical mechanisms for the 
first order transition of our model (for $q>4$), is also very effective in the 
numerical predictions of unknown quantities. Our numerical analysis will be 
completed in a future paper and will concern properties as spontaneous 
magnetization and susceptibility, which are accessible through low temperature 
series expansions \cite{BEG94}, but not through large-$q$ expansions. On the 
other hand, since these two approaches give contradicting results in the case 
of specific heat, a further analysis of the estimates made in ref. \cite{BEG94} 
would be appropriate.
\par Furthermore we will explore more completely the consequences of 
duality, which in the variational approach proposed here appears at a more 
fundamental level. Through the reduced probability distributions and the 
variational parameters having dual partners, it is possible to study more 
completely the link between typical properties of the ordered and disordered 
phases.
\par Our aim will be also to extend the above approach to the important case of 
the 3D Potts model.
\acknowledgments
This research was supported by Ministero dell'Universit\`a e della Ricerca 
Scientifica (40\% and 60\% funds) and Istituto Nazionale di Fisica Nucleare.

%
%

\end{document}